**Machine learning uncovers new cosmological information**

*Large cosmological datasets have been probing the properties of our universe and constraining the parameters of dark matter and dark energy with increasing precision. Deep learning techniques have shown potential to be smarter, and to greatly outperform human-designed statistics.*

**Zoltan Haiman**

The nature of dark energy and dark matter, the dominant components of the present-day universe, remain elusive and represent the most intriguing puzzles in astrophysics. Several cosmology experiments are producing large datasets at unprecedented precision, racing to crack these puzzles. Examples include spectroscopic surveys, measuring three-dimensional galaxy distributions, and imaging surveys, measuring the coherent distortions of apparent galaxy shapes due to gravitational lensing, known as weak lensing. But how much cosmological information is encoded in these datasets? As Dezso Ribli and collaborators [1] demonstrate in this issue of *Nature Astronomy*, potentially a lot more than previously uncovered.

The total information content is simple to pin down in a dataset such as the cosmic microwave background (CMB) anisotropies: the tiny brightness fluctuations across the sky are linear and reflect the Gaussian primordial conditions. In this limit, all information is encoded in the correlation function (or its Fourier counterpart, the power spectrum). Once the mean and the standard deviation of a Gaussian is known, no further information can be gained from any other feature.

By comparison, the three-dimensional mass distribution or the two-dimensional map of the weak lensing distortions represent non-Gaussian random fields, and their total information content is unknown. The coherent lensing distortions in the apparent shapes of distant galaxies can be measured and statistically disentangled from random variations in the intrinsic shapes and orientations of galaxies [2]. This cosmic shear signal has recently been measured using millions of galaxies in the CFHTLenS, DES, and KiDS-450 surveys, and has already yielded competitive constraints on dark matter and dark energy [3-5]. Much larger surveys (Euclid, LSST, WFIRST) are expected to measure the shapes of over a billion galaxies over the whole sky.

The cosmic shear correlation function remains the primary lensing observable. However, on angular scales below a few arcminutes, the shear encodes information from projected small-scale, non-linear density fluctuations, which no longer obey Gaussian distributions. It is increasingly recognized that in this regime, shear maps contain information not accessed by traditional Gaussian statistics. This is illustrated in Figure 1, which shows a "real" simulated lensing map, obtained in a ray-tracing N-body simulation, and a corresponding "fake" version, which is a realization of a Gaussian random field with the same correlation function. The power spectrum or correlation function cannot tell these images apart, but the real map shows conspicuous additional features, such as a larger number of clustered hot-spots.

Many groups have attempted to extract such non-Gaussian information. The most common approach is via statistics that are not determined by the two-point correlation function. These may include higher-order correlation functions or moments of the lensing field (or analogous quantities in Fourier space), or "designer" statistics, such as the abundance of lensing peaks, Minkowski functionals, or wavelets. Other promising ideas include transforming the lensing field into a Gaussian, or reconstructing the Gaussian initial conditions. These approaches improve parameter constraints by up to a factor of two in theoretical studies, and typically by 20-50% when applied to real data (lensing peaks have been a prominent example [6-8]). These are significant improvements already, equivalent to scaling up surveys by a factor of two or more in size. However, are these statistics extracting nearly all of the information from these data, or is there much more? Recent hints suggest that machine learning algorithms could find more.

Artificial neural networks are inspired by natural neural networks, in which a series of nodes ("neurons") are connected to each other in a network. The neurons are typically arranged in layers, perform non-linear operations involving a set of adjustable weights, and transmit the information to the next layer. When used for pattern recognition, the input layer may be an image (a lensing map in our case), and the final output layer may be reduced to just a few numbers (such as cosmological parameter guesses). The weights are adjusted by repeatedly feeding the network with input examples and minimizing a loss function (measuring how outputs differ from known input parameters). This process is called ``training" or ``learning". Deep convolutional neural networks (CNNs) have many layers, some of which convolve their input data with kernels defined by the weights. CNNs yield excellent results for natural images (e.g. telling cats and dogs apart with high fidelity [9]).

Neural networks have long been successfully used in Astronomy, such as in source detection and classification, light curve analysis, or photometric redshift estimation. In the past two years, their utility in cosmological parameter estimation began to be explored, with the goal of eventually applying them to real data . A CNN trained on simulated 3D matter density fields was shown to be able to predict the mean matter density ($\Omega$) and fluctuation amplitude ($\sigma_8$) with a smaller scatter than possible with the power spectrum on the same data [10]. Two groups have subsequently trained CNNs on suites of mock lensing maps [11-12], covering a wide range of cosmologies, and have found that the joint ($\Omega,\sigma_8$) confidence region shrunk by up to a factor of four compared to using the power spectrum.

Ribli et al. have followed up on the last study [12], working with the same set of publicly available (columbialensing.org) simulated lensing maps. They achieved two major results. First, they improved the architecture of the network, which has significantly further reduced both the scatter and the bias in the parameter estimates. Second, they were able to interpret the source of the additional information. By examining the weights in the trained network, they discovered that the network has learned to use the shapes of lensing peaks -- particularly their gradients. They proposed a new statistic (the number of peaks with a given gradient), which out-performs all previous statistics (there was a notable earlier hint [13] that peak shapes are useful).

This latter result is especially significant. The "black box" aspect of neural networks is a common concern and the butt of many jokes in hallway conversations. A commercial firm may ultimately not care how a network distinguishes cats from dogs, as long as it is proven to do it accurately. However, suppose a neural network definitely claims to have discovered, for the first time, evidence that the dark energy equation of state w differs from -1.  Normally, this would definitively rule out finite vacuum energy density as the dark energy. However, we would surely not trust this result from a neural network, unless we understand the physical origin of the information.  Ribli et al. have overcome this concern, by being able to pinpoint the information that the network has uncovered.  Apparently, the shapes of peaks such as the hotspots seen in Figure 1, are sensitive probes of cosmology.

The authors' study is based on idealized, dark-matter only simulations, which leaves significant caveats. Will networks be equally effective on noisy data under realistic survey conditions [14]? Can the impact of baryonic physics on small-scales be mitigated?  The neural network must also avoid any type of over-fitting.  And, finally, if a network is so clever, it may be also be more easily tripped up by systematic errors in real data, delivering biased results.  These questions will surely be addressed in forthcoming research.  However, at this stage, there is no clear showstopper: could machines beat us to the punch and be the first to reveal the true nature of dark energy?

*Zoltan Haiman is at the Department of Astronomy, Columbia University, 1326 Pupin Physics Laboratories, Mailcode 5246, 550 West 120th Street, New York, NY, 10027, U.S.A.*
*e-mail: zoltan@astro.columbia.edu*

| lensing by cosmic structures | mock Gaussian equivalent |
|---|---|
| 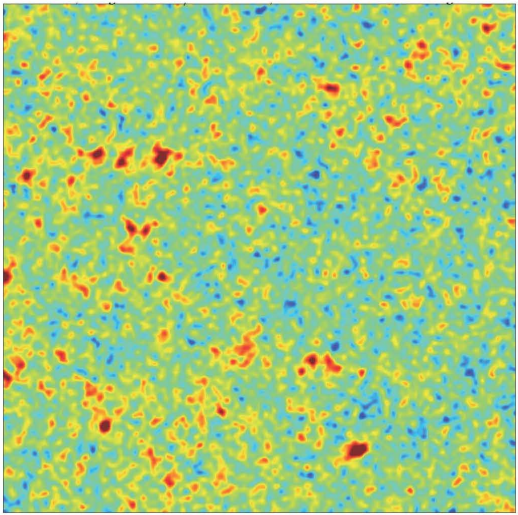 | 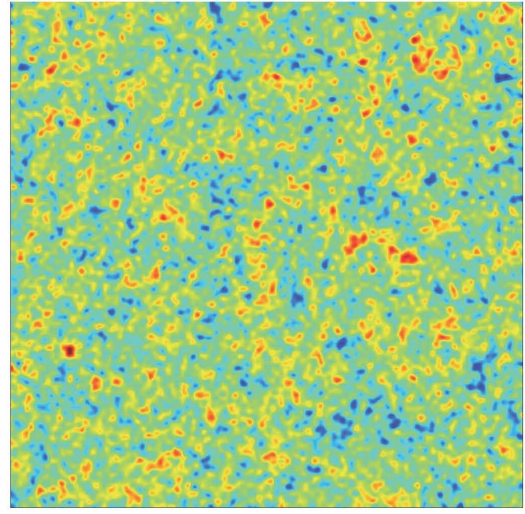 |

**Figure 1 | Comparison between a mock lensing map and a random Gaussian field**. The mock lensing map is of a 3.5 x 3.5 degree patch of the sky, resulting from the projection of large-scale cosmic structures, obtained in a three-dimensional cosmological ray-tracing simulation (ref. [12]; on the left). The corresponding random Gaussian field (on the right) has the same correlation function as the mock lensing map. The "real" map shows significant additional features, such as an excess of large hot spots (areas of strong lensing shear; in red) towards the left side. Credit: Jose Manuel Zorrilla Matilla / columbialensing.org.